\documentclass[pra,twocolumn,nofootinbib,letter]{revtex4}
\usepackage{graphicx}
\usepackage{bm,amsfonts,amsmath,amssymb}
\usepackage{dcolumn}
\usepackage{natbib}
\usepackage%[colorlinks]  %%% sometimes colour links are not seen!
{hyperref}

\hypersetup{
    colorlinks=true,
    linkcolor=blue,
    citecolor=red,
    filecolor=magenta,
    urlcolor=magenta,
}

\newcommand{\Fref}[1]{Fig.~\ref{#1}}

\def\cm{cm$^{-1}$}

\begin{document}

\title{Effective three particle forces in polyvalent atoms}

\author{M. G. Kozlov$^{1,2}$}
\author{M. S. Safronova$^{3,4}$}
\author{S. G. Porsev$^{1,3}$}
\author{I. I. Tupitsyn$^{5}$}

\affiliation{$^1$Petersburg Nuclear Physics Institute, Gatchina
188300, Russia}

\affiliation{$^2$St.~Petersburg Electrotechnical University
``LETI'', Prof. Popov Str. 5, St.~Petersburg, 197376, Russia}

\affiliation{$^{3}$Department of Physics and Astronomy, University
of Delaware, Newark, Delaware 19716, USA}

\affiliation{$^{4}$Joint Quantum Institute, National Institute of
Standards and Technology and the University of Maryland,
Gaithersburg, Maryland 20742, USA}

\affiliation{$^{5}$Department of Physics, St. Petersburg State University,
Ulianovskaya 1, Petrodvorets, St.Petersburg, 198504, Russia}

\date{\today}

\begin{abstract}
We study effective three-particle interactions between valence
electrons, which are induced by the core polarization. Such
interactions are enhanced when valence orbitals have strong overlap
with the outermost core shell, in particular for the systems with
partially filled  $f$-shell. We find that in certain cases the
three-particle contributions are large, affecting the order of
energy levels, and need to be included in high-precision
calculations.
\end{abstract}

\pacs{
 34.80Lx,     % Copied from DFGH12
 31.10.+z,    %
 34.10.+x
}

\maketitle

%%%%%%%%%%%%%%%%%%%%%%%%%%%%%%%%%%%%%%%%%%%%%%%%%%%%%%%%
\section{Introduction}
%%%%%%%%%%%%%%%%%%%%%%%%%%%%%%%%%%%%%%%%%%%%%%%%%%%%%%%%

Accurate prediction of atomic properties is crucial for many
applications, ranging from tests of fundamental physics
\cite{Yb,PruRamPor15} to building ultra-precise atomic clocks
\cite{Sr}. In recent years, atoms and ions with more complicated
electronic structure, including lanthanides and actinides
%became of particular interest
were in the focus of many studies
\cite{lli,hol,er1,Dy-alpha2,lu,cf,SDFS14}. In particular,
highly-charged ions (HCI) with open $nf$-shells have been suggested
for the design of high-precision atomic clocks and the search for the
variation of the fine-structure constant \cite{BDF10,BerDzuFla11b}.
These applications require  accurate predictions of transition
wavelengths and other atomic properties, motivating further
development of high-precision atomic methodologies.

\begin{figure}[htb]
\begin{center}
%\hfill
\includegraphics[scale=0.45]{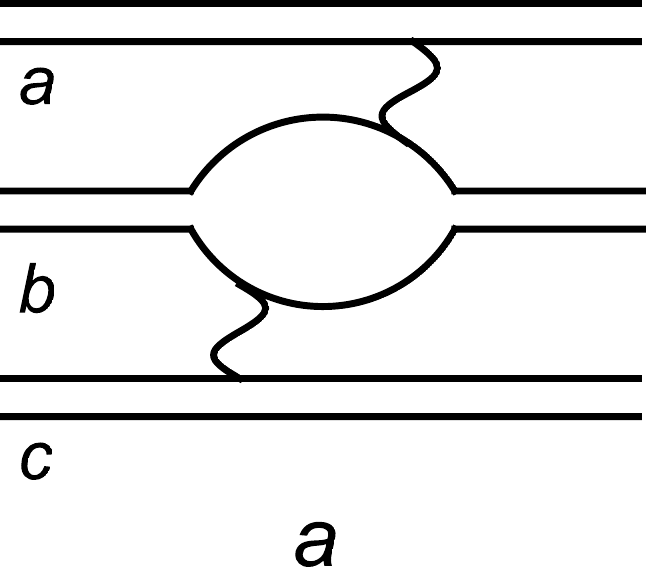}
\hfill
\includegraphics[scale=0.45]{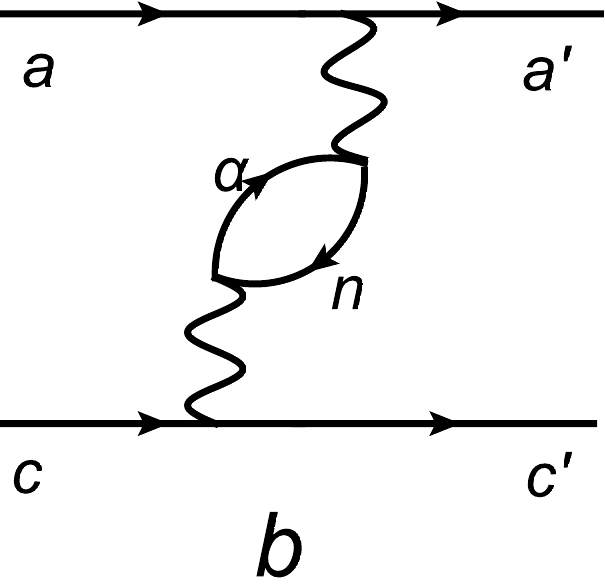}
\hfill
\includegraphics[scale=0.45]{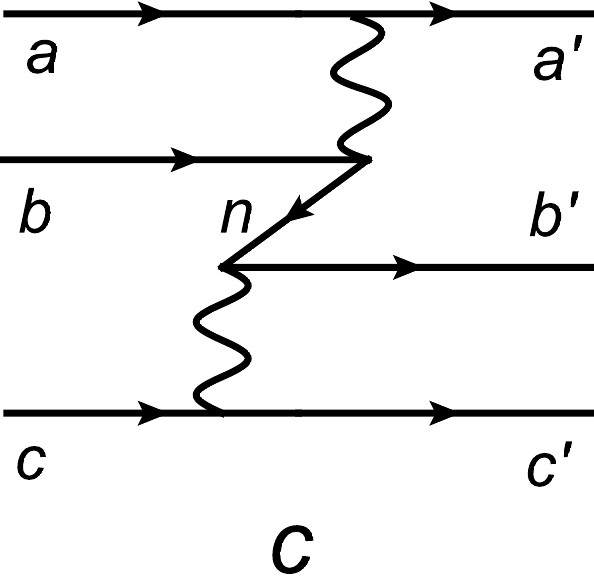}
%\hfill
\end{center}
\caption{Effective interactions. (a) Complex particles: the particle $c$
polarizes the particle $b$, which then interacts with the particle $a$; (b)
Screened interaction: the particle $c$ polarizes the core and interacts
with the particle $a$; (c) The particles $a$, $b$, and $c$ interact through
the excitation of the core.} \label{fig_3n}
\end{figure}

It is well known that three-particle interactions play important
role in nuclear physics. Such interactions arise, for example,
because of the internal structure of the nucleons, see
\Fref{fig_3n}.a. If the nucleon $c$ polarizes the nucleon $b$, then
interaction of the latter with the third nucleon $a$ is modified. In
atomic physics we deal with point-like electrons and such mechanism
of generating effective three-particle interactions is absent.
However, atoms have electronic shell structure and interactions
between valence electrons are modified by the stronger bound core
electrons, which form a kind of inhomogeneous dielectric medium.
This is known as the core-polarization, or the screening effect and is
described by the diagrams of the type of \Fref{fig_3n}.b. The loop
in this diagram includes the sums over  all core states $n$ and all
possible states $\alpha$ above the core. However, some of the states
$\alpha$ can be occupied by valence electrons and should be excluded
due to the Pauli principle. This leads to the diagram
\Fref{fig_3n}.c, which cancels contributions of the states
$\alpha=b,b'$ in the diagram \Fref{fig_3n}.b. Therefore we can say
that three-electron interactions (TEI) between valence electrons
appear because core polarizability depends on the presence of the
valence electrons. Note that TEI are also considered in
condensed matter physics, see, e.~g.\ \cite{MaHo12}.

The diagram \Fref{fig_3n}.c (and its possible permutations) is the
only three-electron diagram in the second order of the many-body
perturbation theory (MBPT) in residual two-electron interaction. In
the case of initial three electron state $(a,b,c)$ and final state
$(a',b',c')$ there are 36 diagrams, which differ by permutations of
these states. This number rapidly grows with the number of valence
electrons and the number of valence configurations, which are
included in the calculation. As a result, the total contributions of
such diagrams for polyvalent atoms may be large.

%------------------------------------------------------------------
%------------------------------------------------------------------

Effective TEI described by the diagram \Fref{fig_3n}.c were
introduced in Ref.\ \cite{DFK96b} within the CI+MBPT approach. This
approach combines the configuration interaction (CI) method for treating
valence correlations with MBPT for core-valence and core-core
correlations. Since then this method was used for calculations of
various properties of polyvalent atoms with several closed core
shells \cite{DFKP98,KP99,KPJ01,SJ02,Dzu05a,Dzu05b,Sav16}. Later, a
CI+AO (all order) method was developed in
\cite{Koz04,SKJJ09,PKST16}. It includes higher-order core valence
correlations by combining configuration interaction and linearized
coupled-cluster approach.

In Ref.\ \cite{DFK96b} neutral Tl was calculated as a three-valence atom and
TEI contribution to the valence energy was found to be very small, on the order of
10 \cm, leading to the omission of TEI contributions in a vast majority of
later calculations. The reason for the suppression of the TEI contribution is
clear from Fig.\ \ref{fig_3n}.c: valence orbitals $b$ and $b'$ typically have
very small overlap with all core orbitals $n$. However, this is not always the
case. When valence $d$, or $f$ shells are filled, they may have relatively
large overlap with the outermost core shell, which in these cases has the same
principal quantum number. In Ref.\ \cite{BFK08} TEI corrections to the
transitions frequencies of Ti$^+$ were found to be from 100 to 200 \cm. The
ground configuration of Ti$^+$ is $3d^2 4s$ and the outermost core shell is
$3p$. The $3d$ and $3p$ shells are not spatially separated and have
significant overlap, resulting in the enhancement of the TEI contributions.

As we noted above, there is significant recent interest in HCI with
optical transitions between the states of configurations with $4f$
and $5f$ electrons \cite{BDF10,BerDzuFla11b}. Two very important
experimental steps toward development of new frequency standards
with these systems and subsequent application to the search for a
possible variation of the fine-structure constant $\alpha$ were
recently completed. First,  predicted $5s$ - $4f$ transitions were
detected in a number of HCI \cite{WLBO15}. Second, sympathetic
cooling of Ar$^{13+}$ with Be$^+$ was demonstrated
\cite{SchVerSch15} paving the way to placing the highly-charged ions
on the same footing as the singly-charged ions such as Al$^+$
currently used for optical atomic clocks \cite{CHKW10}.

Recent work \cite{SDFS14} identified 10 HCI with very narrow optical
transitions, where high precision spectroscopy is possible. All
these ions have atomic cores with 46 electrons $[1s^2\dots 4d^{10}]$
and one to four valence electrons from the $4f$, $5s$, and $5p$
shells. Five ions from this list have three valence electrons:
Ce$^{9+}$, Pr$^{10+}$, Nd$^{11+}$, Sm$^{13+}$, and Eu$^{14+}$. Their
ground configurations are either $5s^2 5p$, or $5s^2 4f$. Pr$^{9+}$
and Nd$^{10+}$ have four valence electrons with ground state
configurations $5s^2 5p^2$ and $5s^2 4f^2$, respectively. We expect
that valence $4f$ orbitals have large overlap with the core shell
$4d$, significantly enhancing three-particle interactions. Since
prediction of accurate transition energies in these highly-charged
ions is crucial for rapid experimental progress, it is important to
evaluate the TEI contributions in these systems, which have been so
far omitted in all relevant HCI calculations.

In this paper we study the role of such effective three-electron interactions
in the spectra of polyvalent atoms and ions. Below we calculate TEI
corrections to transition frequencies of the following ions: Ce$^{9+}$,
Pr$^{9+,10+}$, Nd$^{10+,11+}$, Sm$^{13+}$, and Eu$^{14+}$. We also calculate properties of
the U$^{2+}$ ion as an example of the tetravalent system with the partially filled
$5f$ shell \cite{SaSaSa16}. We find that TEI corrections to the valence
energies are typically of the order of few hundred cm$^{-1}$ in these systems,
but may exceed a thousand cm$^{-1}$. In some cases this is enough to change
the order of low-lying levels significantly affecting theoretical predictions.

\section{Theory}

\begin{figure}[htb]
\begin{center}
\includegraphics[scale=0.45]{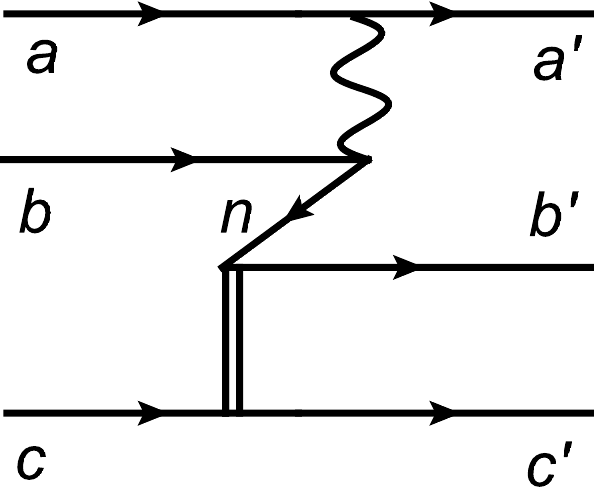}
\end{center}
\caption{The effective three-electron interaction in the coupled-cluster
approximation. Double vertical line corresponds to the two-electron
core-valence cluster amplitude. Such amplitudes are found by solving
standard cluster equations \cite{SKJJ09}. Then, the TEI diagrams are
evaluated using the resulting cluster amplitudes.}
\label{fig_3cluster}
\end{figure}

We use Dirac-Coulomb-Breit Hamiltonian in the no-pair approximation
\cite{Suc80,LKD93}. Low-lying levels of ions are found with the CI+AO method
\cite{SKJJ09}. In this method, the core-valence and core-core correlations are
treated using the linearized coupled-cluster method in the single-double
approximation \cite{BJS91,SJD99} instead of the second order MBPT used in the
CI+MBPT approach. A complete treatment of the TEI at the CI+AO level involves
modification of the TEI diagrams in Fig.\ \ref{fig_3n}.c to the form presented
in Fig.\ \ref{fig_3cluster}, where one Coulomb-Breit interaction is
substituted by the respective cluster core-valence amplitude \cite{Koz04}.
%in the CI+AO method.
However, we find it sufficient to carry out CI+AO calculations of
the wavefunctions and then treat TEI corrections within the second
order MBPT for the systems of interest.

%------------------------------------------------------------------
%------------------------------------------------------------------

Our initial approximation corresponds to the Hartree-Fock potential
of the core, $V^{N_c}$, where $N_c$ is the number of core electrons.
Such approximation completely neglects interactions between valence
electrons and may be too crude for some neutral polyvalent atoms
\cite{PKST16}, but is sufficiently good for HCIs. Next, we form an
effective Hamiltonian for valence electrons:
\begin{equation}
 H_{\rm eff}(E) = H_{\rm FC} + \Sigma(E), \label{Heff1}
\end{equation}
where $H_{\rm FC}$ is the Hamiltonian in the frozen-core
approximation, which includes Coulomb-Breit interactions between
valence electrons and the core potential $V^{N_c}$.

The energy-dependent operator $\Sigma(E)$ accounts for the core polarization
effects, such as in Fig.\ \ref{fig_3n}.b. In the second order of MBPT this
operator is a three-electron operator. In higher orders it is the
$N_v$-electron operator, where $N_v$ is the number of valence electrons (we
assume that $N_v\ge 3$ and $N=N_c+N_v$ is the total number of electrons in the
system). At this stage we neglect three-electron and many-electron
interactions and consider operator $\Sigma$ as a two-electron operator.
Explicit expressions for $\Sigma$ are given in Refs.\ \cite{DFK96b,SKJJ09}.
%%%%%%%%%%%%%%%%%%%%%%
%It is constructed using the second order many-body perturbation
%theory in the CI+MBPT approach \cite{DzuFlaKoz96} or linearized
%coupled cluster single-double (LCCSD) method in the CI+all-order
%approach~\cite{SafKozJoh09}.
%%%%%%%%%%%%%%%%%%%%%%
%These interactions are modified to account
%%%%%%%%%%%%%%%%%%%%%%
We use the Davidson algorithm to find $L$ lowest eigenvalues and
eigenfunctions of the operator $H_{\rm eff}$ (typically $L\sim 10$).

Selection rules for three-electron matrix elements are much weaker
than for two-electron ones and the number of nonzero matrix elements
of the effective Hamiltonian drastically increases. Consequently,
the matrix becomes less sparse. Forming and diagonalizing such
matrix in a complete configurational space is impractically
time-consuming. Instead, we include TEI by forming a small $L\times
L$ matrix using eigenfunctions from the previous stage of the
computation. Diagonalization of this matrix gives us eigenvalues
with TEI corrections. This approach radically reduces the number of
required three-electron diagrams without significant loss of
accuracy.

\section{Results and discussion}

\subsection{In-like and Sn-like HCI with narrow optical transitions}

For In-like Ce$^{9+}$, Pr$^{10+}$, Nd$^{11+}$, Sm$^{13+}$, and
Eu$^{14+}$ ions (49 electrons) and for Sn-like Pr$^{9+}$ and
Nd$^{10+}$ ions (50 electrons) we use the results of previous CI+AO
calculations with Dirac-Coulomb-Breit two-electron effective
Hamiltonian described in detail in Refs.\ \cite{SDFS14a} and
\cite{SDFS14b} respectively. In-like ions were calculated in
\cite{SDFS14a} in two approximations, either as systems with  one,
or three valence electrons. Similarly, Sn-like ions were treated in
\cite{SDFS14b} as systems with two, or four valence electrons.
Calculations with three and four valence electrons include
correlations more completely and are expected to be more accurate.
On the other hand, in these approximations we need to include TEI
contributions. In this work, we use eigenfunctions obtained in
\cite{SDFS14,SDFS14a,SDFS14b} and add TEI corrections as discussed
above.

%------------------------------------------------------------------
\begin{figure}[htb]
\begin{center}
\hfill
\begin{minipage}[c] {30mm}
\includegraphics[scale=0.45]{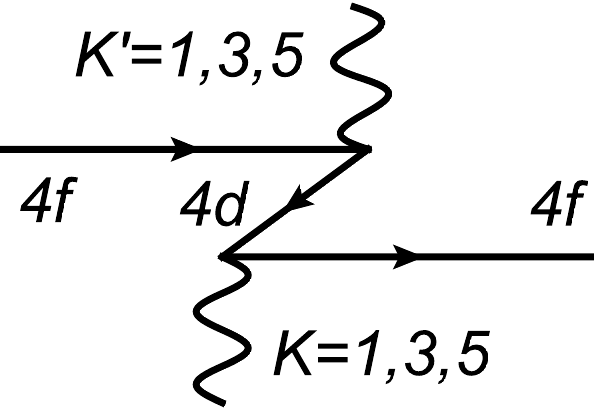}
\end{minipage}
\hfill
\begin{minipage}[c] {30mm}
\includegraphics[scale=0.45]{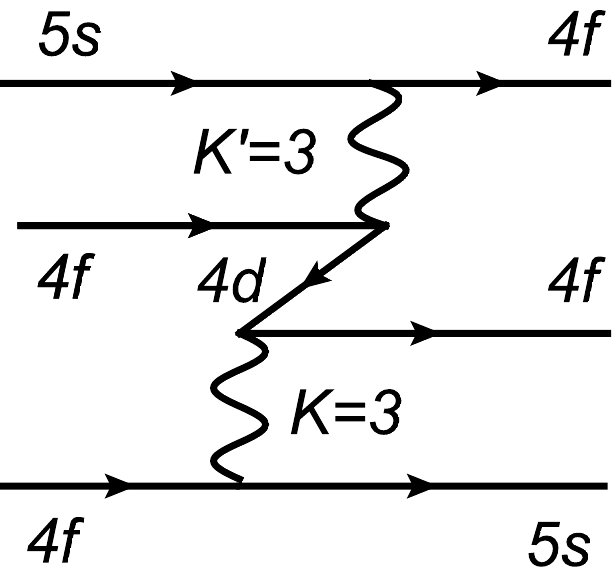}
\end{minipage}
\hfill
 $\,$
\end{center}
\caption{TEI for the case of the valence $4f$ electron and the $4d$ core
shell. Left panel: selection rules for the $4d$-$4f$ vertex require odd
multipoles $K,K'$ for the Coulomb interaction. Right panel: example of the
nonzero diagram for the configuration $5s4f^2$. Selection rules for the $5s^2
4f$ and $4f^3$ configurations require even multipoles and prohibit vertexes
with $4d$ core shell electrons; thus, for these configuration the $4d$
contribution to TEI diagrams vanishes.}
 \label{fig_3example}
\end{figure}
%------------------------------------------------------------------

Trivalent ions considered in this work  have the following low-lying
valence configurations with $4f$ electrons: $5s^2 4f$, $5s4f^2$, and
$4f^3$. Fig.\ \ref{fig_3example} illustrates that the contribution
from the uppermost $4d$ core shell in TEI diagrams vanishes for the
$5s^2 4f$ and $4f^3$ configurations. Therefore, we can expect large
TEI corrections only for the $5s4f^2$ configuration. In Ce$^{9+}$,
Pr$^{10+}$, and Nd$^{11+}$ this configuration lies very high and is
not of interest  to clock applications. Only in Sm$^{13+}$ this
configuration is within the optical range transition from the ground
configuration $5s^2 4f$. In Eu$^{14+}$ the $5s4f^2$ configuration
becomes the ground one. Consequently, the TEI corrections to the
energies of the low-lying levels of Ce$^{9+}$, Pr$^{10+}$, and
Nd$^{11+}$ are rather small, but become much larger for Sm$^{13+}$
and Eu$^{14+}$. For the former group of ions these corrections are
on the order of 100 \cm or less, but for the latter group they
exceed 500 \cm.

%------------------------------------------------------------------
\begin{figure*}[htb]
\begin{center}
%\hfill
\includegraphics[scale=0.5]{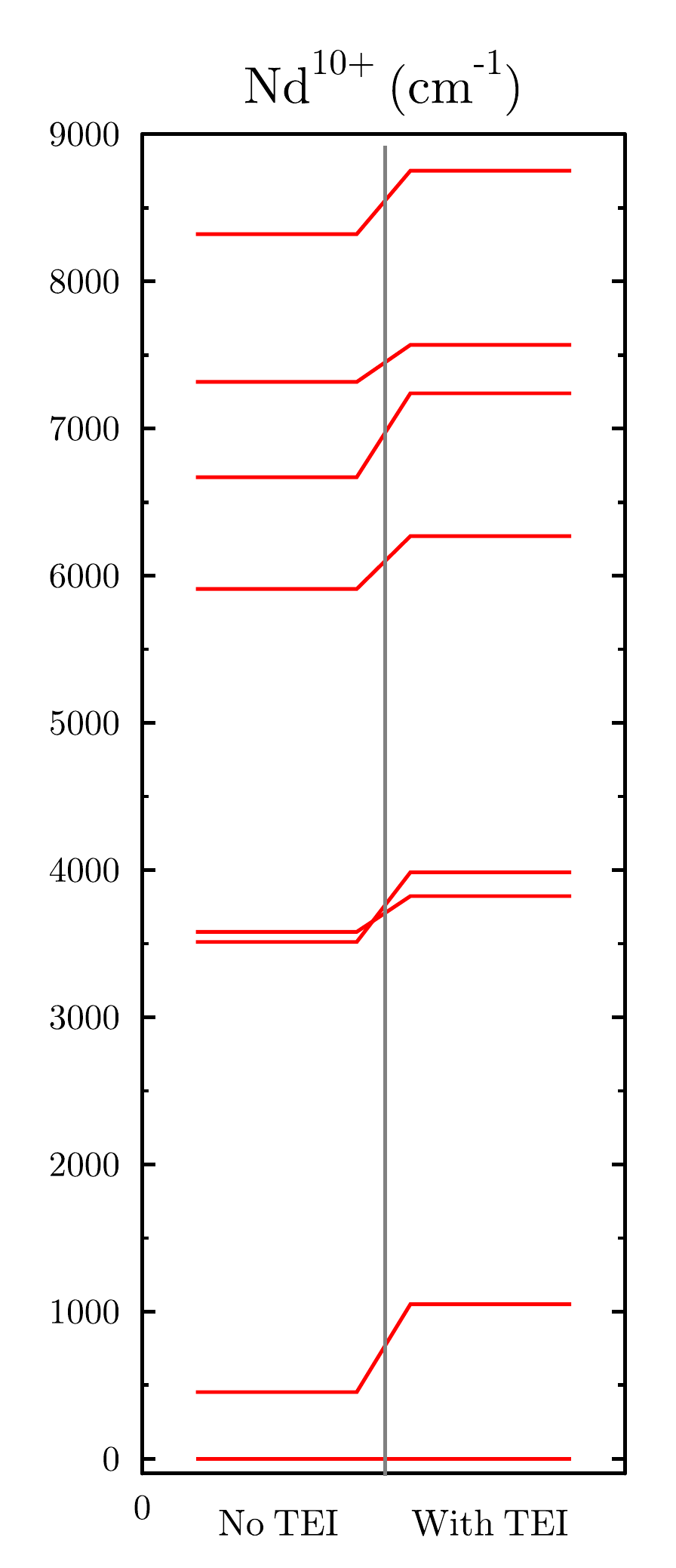}
\hfill
\includegraphics[scale=0.5]{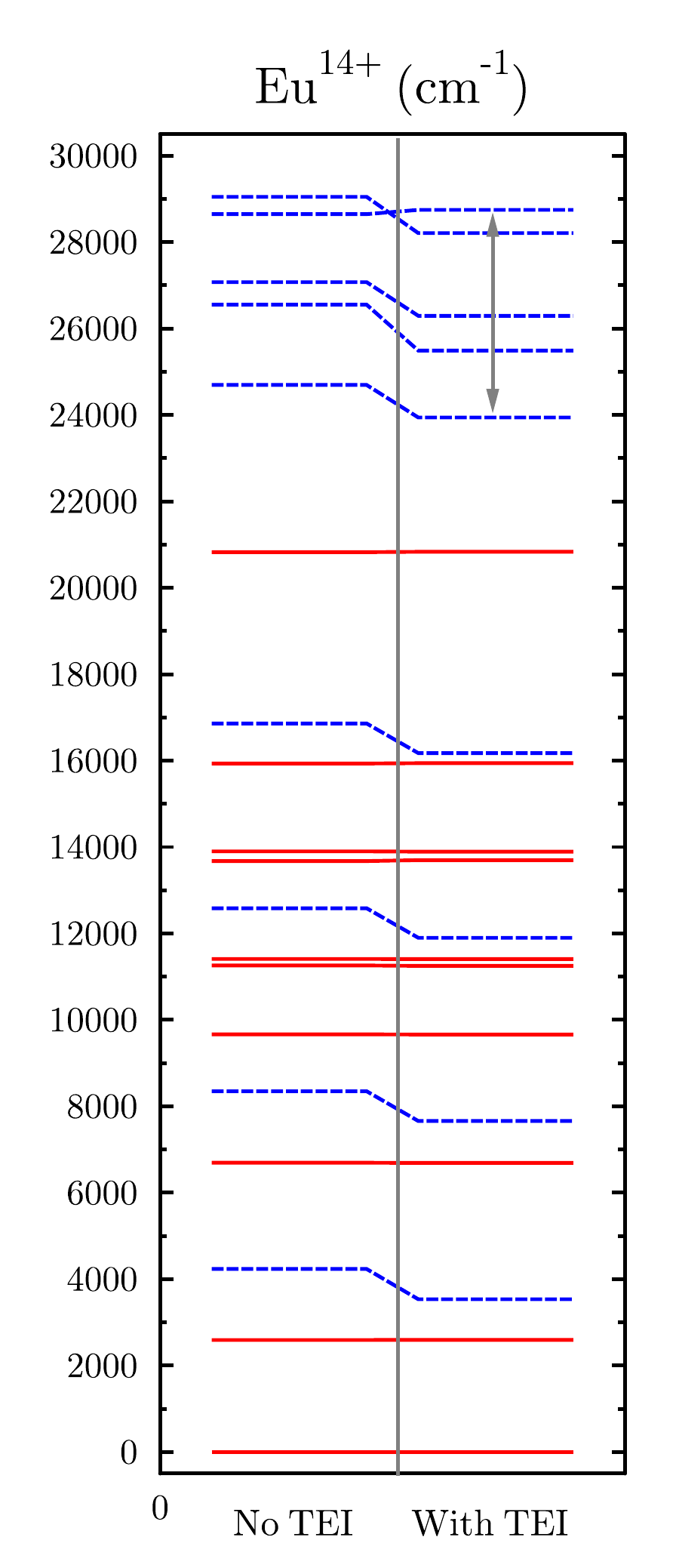}
\hfill
\includegraphics[scale=0.5]{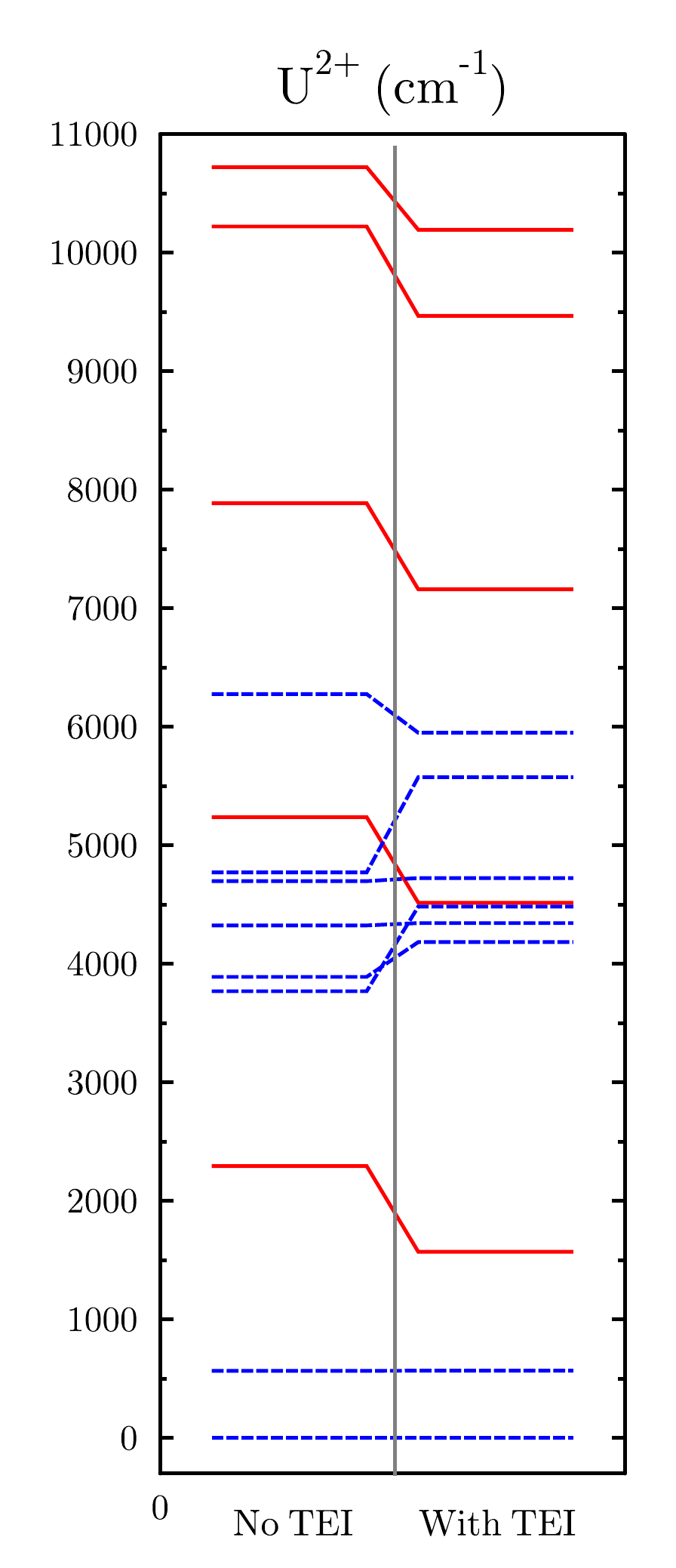}
%\hfill
\end{center}
\caption{(color online) Level diagrams of Nd$^{10+}$, Eu$^{14+}$, and U$^{2+}$
ions with and without TEI corrections. Solid red lines -- levels of even
parity; dashed blue -- odd parity. Vertical arrow in the central panel shows
two strongly interacting levels.}
 \label{fig_U2spectrum}
\end{figure*}
%------------------------------------------------------------------

Results of our calculations for HCI with three valence electrons are
presented in Table \ref{Tab_3e_HCI}. The spectrum of Eu$^{14+}$ is also
shown in the central panel of Fig.\ \ref{fig_U2spectrum}. The TEI
corrections shift levels of the odd parity down by approximately 500
\cm, with the only exception of one level at the top of the
plot. For this level there is large non-diagonal TEI interaction
with the lower level of the same $J$ and parity. This interaction
is shown by the vertical arrow.

\begin{table}[htb]
\caption{Calculated low-lying levels of Ce$^{9+}$, Pr$^{10+}$,
Nd$^{11+}$, Sm$^{13+}$, and Eu$^{14+}$. Column 4 lists excitation
energies in the CI+AO approximation from Ref.\ \cite{SDFS14a}. TEI
corrections to the valence energy and respective shifts relative to
the ground state are given in columns 5 and 6. Last column
presents final calculated spectra. All values are in \cm.}
 \label{Tab_3e_HCI}
 \begin{tabular}{cccrrrr}
 \hline\hline
 \multicolumn{1}{c}{Ion}
 &
 \multicolumn{1}{c}{Config.}
 & \multicolumn{1}{c}{$J$}
 & \multicolumn{1}{c}{CI+AO}
 & \multicolumn{1}{c}{TEI}
 & \multicolumn{1}{c}{$\Delta_\mathrm{TEI}$}
 & \multicolumn{1}{c}{Total}
 \\
 \hline\\[-3mm]
 Ce$^{9+}$ & $5s^2 5p$ & $\tfrac12$  &     0 &$  171$&$    0$&     0 \\[0.5mm]
           & $5s^2 5p$ & $\tfrac32$  & 33436 &$  177$&$    6$& 33442 \\[0.5mm]
           & $5s^2 4f$ & $\tfrac52$  & 55694 &$  126$&$  -45$& 55649 \\[0.5mm]
           & $5s^2 4f$ & $\tfrac72$  & 58239 &$  121$&$  -50$& 58189 \\[0.5mm]
 \hline\\[-3mm]
 Pr$^{10+}$& $5s^2 5p$ & $\tfrac12$  &     0 &$  183$&$    0$&     0 \\[0.5mm]
           & $5s^2 4f$ & $\tfrac52$  &  4496 &$  147$&$  -36$&  4460 \\[0.5mm]
           & $5s^2 4f$ & $\tfrac72$  &  7817 &$  141$&$  -42$&  7776 \\[0.5mm]
           & $5s^2 5p$ & $\tfrac32$  & 39127 &$  190$&$    7$& 39134 \\[0.5mm]
 \hline\\[-3mm]
 Nd$^{11+}$& $5s^2 4f$ & $\tfrac52$  &     0 &$  167$&$    0$&     0 \\[0.5mm]
           & $5s^2 4f$ & $\tfrac72$  &  4173 &$  160$&$   -7$&  4167 \\[0.5mm]
           & $5s^2 5p$ & $\tfrac12$  & 52578 &$  198$&$   31$& 52609 \\[0.5mm]
           & $5s^2 5p$ & $\tfrac32$  & 97945 &$  205$&$   39$& 97984 \\[0.5mm]
 \hline\\[-3mm]
 Sm$^{13+}$& $5s^2 4f$ & $\tfrac52$  &     0 &$  205$&$    0$&     0 \\[0.5mm]
           & $5s^2 4f$ & $\tfrac72$  &  6165 &$  197$&$   -8$&  6157 \\[0.5mm]
           & $5s 4f^2$ &$\tfrac{11}2$& 22521 &$  530$&$  326$& 22847 \\[0.5mm]
           & $5s 4f^2$ &$\tfrac32   $& 24774 &$  531$&$  326$& 25100 \\[0.5mm]
           & $5s 4f^2$ &$\tfrac{13}2$& 28135 &$  527$&$  322$& 28458 \\[0.5mm]
           & $5s 4f^2$ & $\tfrac52$  & 31470 &$  528$&$  324$& 31794 \\[0.5mm]
 \hline\\[-3mm]
 Eu$^{14+}$& $5s 4f^2$ & $\tfrac72$  &     0 &$  574$&$    0$&     0 \\[0.5mm]
           & $5s 4f^2$ & $\tfrac92$  &  2592 &$  575$&$    1$&  2593 \\[0.5mm]
           & $   4f^3$ & $\tfrac72$  &  4235 &$ -126$&$ -700$&  3535 \\[0.5mm]
           & $5s 4f^2$ &$\tfrac{11}2$&  6694 &$  569$&$   -4$&  6690 \\[0.5mm]
           & $   4f^3$ &$\tfrac{11}2$&  8348 &$ -115$&$ -689$&  7659 \\[0.5mm]
           & $5s 4f^2$ &$\tfrac32   $&  9664 &$  571$&$   -3$&  9662 \\[0.5mm]
           & $5s 4f^2$ &$\tfrac{13}2$& 11259 &$  565$&$   -9$& 11250 \\[0.5mm]
           & $5s 4f^2$ & $\tfrac52$  & 11410 &$  570$&$   -3$& 11407 \\[0.5mm]
           & $   4f^3$ &$\tfrac{11}2$& 12583 &$ -110$&$ -684$& 11900 \\[0.5mm]
 \hline\hline
 \end{tabular}
\end{table}

\begin{table}[htb]
\caption{Calculated low-lying levels of Pr$^{9+}$ and Nd$^{10+}$
(\cm). Notations are the same as in  Table \ref{Tab_3e_HCI}.}
 \label{Tab_4e_HCI}
 \begin{tabular}{cccrrrr}
 \hline\hline
 \multicolumn{1}{c}{Ion}
 &
 \multicolumn{1}{c}{Config.}
 & \multicolumn{1}{c}{$\,\,J\,$}
 & \multicolumn{1}{c}{CI+AO}
 & \multicolumn{1}{c}{TEI}
 & \multicolumn{1}{c}{$\Delta_\mathrm{TEI}$}
 & \multicolumn{1}{c}{Total}
 \\
 \hline\\[-3mm]
 Pr$^{9+}$ & $5s^2 5p^2$ & 0 &     0 &$  571$&$    0$&     0 \\[0.5mm]
           & $5s^2 5p4f$ & 3 & 22918 &$  544$&$  -28$& 22891 \\[0.5mm]
           & $5s^2 5p4f$ & 2 & 25022 &$  874$&$  303$& 25325 \\[0.5mm]
           & $5s^2 5p4f$ & 3 & 28023 &$  692$&$  121$& 28143 \\[0.5mm]
           & $5s^2 5p^2$ & 1 & 28422 &$  606$&$   34$& 28456 \\[0.5mm]
           & $5s^2 5p4f$ & 4 & 30370 &$  396$&$ -175$& 30195 \\[0.5mm]
           & $5s^2 5p^2$ & 2 & 36459 &$  720$&$  149$& 36607 \\[0.5mm]
           & $5s^2 5p4f$ & 3 & 56234 &$  869$&$  298$& 56532 \\[0.5mm]
 \hline\\[-3mm]
 Nd$^{10+}$& $5s^2 5p4f$ & 3 &     0 &$  534$&$    0$&     0 \\[0.5mm]
           & $5s^2 4f^2$ & 4 &   454 &$ 1115$&$  581$&  1035 \\[0.5mm]
           & $5s^2 4f^2$ & 2 &  3580 &$  828$&$  293$&  3873 \\[0.5mm]
           & $5s^2 4f^2$ & 5 &  3512 &$ 1104$&$  569$&  4081 \\[0.5mm]
           & $5s^2 5p4f$ & 3 &  5910 &$  772$&$  238$&  6147 \\[0.5mm]
           & $5s^2 4f^2$ & 6 &  6669 &$ 1093$&$  559$&  7228 \\[0.5mm]
           & $5s^2 5p4f$ & 4 &  7316 &$  698$&$  164$&  7480 \\[0.5mm]
           & $5s^2 4f^2$ & 2 &  8320 &$  975$&$  441$&  8761 \\[0.5mm]
 \hline\hline
 \end{tabular}
\end{table}

Tetravalent Pr$^{9+}$ and Nd$^{10+}$ ions have low-lying $5s^2
5p^2$, $5s^2 5p4f$, and $5s^2 4f^2$ configurations. There are no contributions of the
uppermost core shell $4d$ to the TEI diagrams for the pure $5s^2 5p^2$
configuration. On the other hand, the configuration interaction for these ions is
stronger than for three electron ions and the $4d$ shell contributes even to those
levels, which nominally belong to the $5s^2 5p^2$ configuration. Moreover, the
number of permutations of the TEI diagrams for four-electron ions is larger
leading to an additional enhancement of the TEI corrections. Our results are
presented in Table \ref{Tab_4e_HCI}. Spectrum of Nd$^{10+}$ is also shown in
Fig.\ \ref{fig_U2spectrum}. We see that TEI corrections for all configurations
are positive and large, on the order of 1000 \cm. Respective energy shifts
relative to the ground state are significantly smaller, about 600 \cm, or
less. We conclude that the size of TEI corrections for Pr$^{9+}$ and
Nd$^{10+}$ is not so sensitive to the leading configuration and, therefore, is
less predictable based on the selection rule arguments, since it is
significantly affected by the configuration interaction.

\subsection{The U$^{2+}$ ion}

\begin{table}[htb]
\caption{Calculated levels of U$^{2+}$ (\cm). Eight lowest levels of
each parity are listed. Notations are the same as in Table \ref{Tab_3e_HCI}.}
 \label{Tab_U2}
 \begin{tabular}{cccrrrr}
 \hline\hline
 \multicolumn{1}{c}{Ion}
 &
 \multicolumn{1}{c}{Config.}
 & \multicolumn{1}{c}{$\,\,J\,$}
 & \multicolumn{1}{c}{CI+AO}
 & \multicolumn{1}{c}{TEI}
 & \multicolumn{1}{c}{$\Delta_\mathrm{TEI}$}
 & \multicolumn{1}{c}{Total}
 \\
 \hline\\[-3mm]
  U$^{2+}$ & $6d   5f^3$ & 6 &     0 &$  679$&$    0$&     0 \\[0.5mm]
           & $6d   5f^3$ & 5 &   567 &$  680$&$  -13$&   568 \\[0.5mm]
           & $5f^4     $ & 4 &  2294 &$  -45$&$ -724$&  1571 \\[0.5mm]
           & $6d   5f^3$ & 3 &  3890 &$  972$&$  130$&  4184 \\[0.5mm]
           & $5f^3 7s  $ & 7 &  4324 &$  698$&$   80$&  4344 \\[0.5mm]
           & $6d   5f^3$ & 4 &  3769 &$ 1393$&$  333$&  4483 \\[0.5mm]
           & $5f^4     $ & 5 &  5238 &$  -44$&$ -723$&  4515 \\[0.5mm]
           & $6d   5f^3$ & 6 &  4698 &$  704$&$ -181$&  4724 \\[0.5mm]
           & $5f^3 7s  $ & 5 &  4771 &$ 1482$&$  145$&  5575 \\[0.5mm]
           & $6d   5f^3$ & 4 &  6276 &$  352$&$  304$&  5949 \\[0.5mm]
%\hline\\[-3mm]
           & $5f^4     $ & 6 &  7886 &$  -48$&$ -727$&  7159 \\[0.5mm]
           & $5f^4     $ & 7 & 10221 &$  -76$&$ -755$&  9466 \\[0.5mm]
           & $5f^4     $ & 4 & 10722 &$  149$&$ -529$& 10192 \\[0.5mm]
           & $5f^4     $ & 3 & 11677 &$  369$&$ -309$& 11368 \\[0.5mm]
           & $5f^4     $ & 8 & 12345 &$ -115$&$ -794$& 11551 \\[0.5mm]
           & $5f^4     $ & 3 & 12660 &$  355$&$ -324$& 12336 \\[0.5mm]
 \hline\hline
 \end{tabular}
\end{table}

In this section we consider U$^{2+}$ as an example of an ion with the partially
filled $5f$ shell. This ion has 4 valence electrons and the $[1s^2 \dots
5d^{10} 6s^2 6p^6]$ core. Low-lying configurations include $5f^3 6d$, $5f^3
7s$, and $5f^4$. Here, two valence orbitals have large overlap with the core:
$5f$ overlaps with the $5d$ shell, while $6d$ overlaps with the $6p$ shell. As a
result, the TEI corrections are very large for the $5f^3 6d$ and $5f^3 7s$
configurations. For the $5f^4$ configuration selection rules for multipoles
suppress the TEI corrections. We use results from Ref.\ \cite{SaSaSa16} as a
starting point for our calculation. In Table \ref{Tab_U2} we present the
calculated spectrum of U$^{2+}$ from Ref.\ \cite{SaSaSa16} and our TEI
corrections to the energies. Both spectra are also shown in the right panel of
Fig.\ \ref{fig_U2spectrum}.

We see that TEI corrections in U$^{2+}$ are large and significantly
differ even for levels of the same configuration. This can be
explained by the large number of diagrams for  the four-electron system
which can either add coherently or  cancel each
other. As expected, TEI corrections for the levels
of the $5f^4$  configuration are several times smaller than for two
other configurations due to selection rules.

The U$^{2+}$ ion has the very dense spectrum with a typical level spacing of few
hundred \cm\ even near the ground state. This is much smaller than the average TEI
correction. Moreover, the dispersion of TEI corrections is also larger than
the typical level spacing. Thus, it is not surprising that the order of levels
appears to be significantly different when TEI corrections are taken into
account (see the right panel of Fig.\ \ref{fig_U2spectrum}). We note, however,
that TEI corrections are insufficient to significantly improve agreement
between our theory and the experiment for U$^{2+}$.

\subsection{Accuracy analysis}

Let us briefly discuss how accurately we account for TEI interactions.
Potentially there are three sources of errors:\\
%
%\begin{itemize}
%\item
\textit{(1) Incompleteness of the one-electron basis set.} It is clear from
\Fref{fig_3n}.c that in TEI diagrams we do not sum over intermediate states
(the only sum is over core states), so there is no error associated with the
final basis set.
\\
%\item
\textit{(2) The truncation of the contributions from the subdominant
configurations.} We neglect small contributions to the eigenfunctions when
calculating TEI corrections. Typically, the configurational mixing accounts for 10\%
correction to the binding energy. Main part of this corrections comes from
the small number of leading configurations, which we take into account. We
estimate the neglected part of this correlation correction to be on the order
of 2~--~3\% of the largest TEI correction.
\\
%\item
\textit{(3) High-order corrections to TEI diagrams.} We calculate TEI
corrections within the second order MBPT, \Fref{fig_3n}.c, instead of using
more accurate expression \Fref{fig_3cluster}. Higher-order terms typically
give 5~--~10\% corrections to the second order diagrams. As long as the
cluster amplitudes in the diagram \Fref{fig_3cluster} are the same as in
two-electron valence diagrams, we can expect similar size of the high-order
corrections here, i.e.\  5~--~10\%.
%\end{itemize}

We conclude that our error for the TEI contribution can be up to 10\%.
According to this estimate we can assign the TEI error bar to be ~50~\cm\ for
three-electron ions from Table \ref{Tab_3e_HCI} and about 100~\cm\ for
four-electron ions from Table \ref{Tab_4e_HCI}. For U$^{2+}$ both CI and
high-order errors are the largest. We can assume here a conservative error bar
of 200 \cm. All these error bars for TEI corrections are smaller than the
total theoretical errors, so they do not affect the overall accuracy of the
theory.

%This seems reasonable. If TEI higher-orders scale the same way as the total
%correlation correction , higher-order error may be estimated from difference
%of the  MBPT and All-order results from previous papers where we had these
%ions, HCI paper definitely had the data. 5-10% seems reasonable anyway.

\section{Conclusions}

We calculated corrections to the energies of several heavy
polyvalent ions from the effective three-electron interactions
induced by the core polarization. We find that these corrections may
be on the order of 1000 \cm for systems with partially filled $4f$,
or $5f$ shells. Atoms and ions with the partly filled $f$-shell usually
have very dense spectrum and TEI corrections can change the
predicted order of energy levels. Large TEI diagrams obey specific
selection rules. For some configurations these selection rules
cannot be satisfied suppressing the TEI corrections for levels of
such configurations.

The number of TEI diagrams rapidly grows with the number of valence
electrons and Hamiltonian matrix becomes less sparse. This makes it
very difficult to account for TEI corrections accurately when they
become large. Here we used relatively simple approximation when we
calculated TEI corrections only in a small subspace spanned by lower
eigenvectors of the unperturbed problem. This method works for the
eigenvalues, but may be insufficient for other observables.

Finally, we  note that ions considered here are sufficiently heavy for QED
corrections to be important. In fact, QED corrections appear to be of the same
order as TEI corrections considered here. Therefore, accurate calculations
have to account for both types of corrections. However, an accurate treatment
of QED corrections in many-electron systems is highly non-trivial
\cite{FlaGin05,STY13,STY15,GiBe16} and this topic is studied elsewhere
\cite{TKSSD16}.

\acknowledgements

This work is partly supported by Russian Foundation for Basic
Research Grant No.\ 14-02-00241 and by U.S. NSF Grants No.\
PHY-1404156 and No. \ PHY-1520993. One of us (IT) acknowledges
support from Grant SPbSU No.\ 11.38.261.2014.

%\bibliographystyle{apsrev}
%\bibliography{../Bib/full,../Bib/my_ref_w,../Bib/my_notes}
%%\bibliography{./3e}
%\end{document}
%%%%%%%%%%%%%%%%%%%%%%%%%%%%%%%%%%%%%%%%%%%%%%%%%%%%
%%%%%%%%%%%%%%%%%%%%%%%%%%%%%%%%%%%%%%%%%%%%%%%%%%%%

\end{document}